\documentclass{emulateapj}
\usepackage{epsfig}
\usepackage{amsmath}
\usepackage{natbib}

\usepackage{lineno}

\def \teff {$T_{\rm eff}$}

\def \logg {$\rm log(g)$}
\def \ttau {$T$-$\tau$}
\def \qrad {$Q_{\rm rad}$}

\def \pturbopgas {$P_{\rm turb} / P_{\rm gas}$}

\begin{document}

%\linenumbers

\title{Comparing the Effect of Radiative Transfer Schemes on Convection Simulations}
\author{Joel D. Tanner, Sarbani Basu \& Pierre Demarque}
\affil{Astronomy Department, Yale University, P.O. BOX 208101, New Haven, CT 06520-8101}
%\date{\today}

\begin{abstract}
We examine the effect of different radiative transfer schemes on the properties of 3D simulations of near-surface stellar convection in the superadiabatic layer, where energy transport transitions from fully convective to fully radiative. We employ two radiative transfer schemes that fundamentally differ in the way they cover the 3D domain. The first solver approximates domain coverage with moments, while the second solver samples the 3D domain with ray integrations. By comparing simulations that differ only in their respective radiative transfer methods, we are able to isolate the effect that radiative efficiency has on the structure of the superadiabatic layer. We find the simulations to be in good general agreement, but they show distinct differences in the thermal structure in the superadiabatic layer and atmosphere.
\end{abstract}

\section{Introduction}

Techniques for constructing 1D stellar models continually evolve in order to encompass more realistic physics.  A stellar structure model is computed with a given set of physical assumptions and boundary conditions.  Stellar evolution models with convective envelopes require an approximation for convection to define the structure within the convection zone.  Current stellar models with convective envelopes are not accurate near the surface.  This is largely due to the lack of a proper description for inefficient convection. 

The Mixing Length Theory \citep[MLT;][]{1958ZA.....46..108B} is the most commonly used approximation for convection in 1D stellar models.  MLT represents the range of turbulent eddy-sizes with a single scale length that is proportional to the pressure scale height.  The so-called ``mixing length" is $l=\alpha H_{\rm P}$, where $\alpha$ is a free parameter.  This prescription is able to accurately model the stellar structure where convection is efficient and the temperature gradient is nearly adiabatic.  In the superadiabatic layer (SAL) near the surface, however, the energy transport transitions from convective to radiative and MLT breaks down.  Note that some MLT treatments \citep[see e.g.,][]{2010ApJ...710.1619A} include an additional free parameter that sets the ratio between the eddy size and the distance it travels. Other approximations for convection exist, such as the treatment of \citet{1991ApJ...370..295C}, but it, too, contains a length scale which has to be modeled.  In this region of inefficient convection and convective overshoot we are unable to accurately model stellar structure.  

One of the major limitations of using MLT or similar treatments of convection in stellar models is that it effectively renders the stellar radius to be a free parameter.  In 1D models with a given set of physics, the mixing length parameter $\alpha$ determines the specific entropy of the convection zone, which in turn sets the radius of the model.  In addition to the approximation for convection, stellar models require an atmospheric boundary condition.  Many stellar evolution codes force the atmospheric structure to be the Eddington {\ttau} relation, although other model atmospheres can be used as well.  In the case of the Eddington {\ttau} relation, the atmosphere is purely radiative, and does not include the effect of convective overshoot.  This, too, contributes to the inaccuracy of near-surface layers in stellar models.

A promising strategy for improving our understanding of inefficient convection is to use 3D radiation hydrodynamic (RHD) large eddy simulations (LES) to calculate a more realistic stratification.  Pioneered by \citet{1982A&A...107....1N,1985SoPh..100..209N}, such simulations account for both the realistic transfer of energy through convective gas dynamics and radiative transfer in the SAL where energy transport changes from convection to radiation.  Simulations provide a realistic stratification that self-consistently couples the convective envelope to the radiative atmosphere. Simulating convection has been feasible for several decades, and a number of groups \citep[e.g.][etc.]{1989ApJ...336.1022C,1989ApJ...342L..95S,1998ApJ...499..914S,2000SoPh..192...91S,1991ApJ...370..282C,1994ApJS...93..309P,1995ApJ...442..422K,2003MNRAS.340..923R,2004MNRAS.347.1208R,2005MNRAS.362.1031R,2007ASPC..362..306J,2011ASPC..448..855F}  have performed 3D simulations of convection.  Simulations have been used to study 3D penetrative convection both above and below convective envelopes \citep[e.g.][]{1994A&A...281L..73S,1995A&A...295..703S,1995A&A...293..127M} and have also been compared with 1D stellar and atmosphere models \citep[e.g.][]{1990A&A...228..155N,1996ApJ...461..499K,1996A&A...313..497F,1997ApJ...480..395A,2000ASPC..203..362N,2008IAUS..252..253P,2011A&A...526A.100P,2011sf2a.conf..235P,2011sf2a.conf..407P}.

Several codes are currently being used by various groups to simulate near surface convection.  These codes include STAGGER \citep{Nordlund1995,1996JGR...10113445G,1998ApJ...499..914S,2011A&A...529A.158H}, CO5BOLD \citep{2002AN....323..213F,2012JCoPh.231..919F}, MURaM \citep{2005A&A...429..335V}, ANTARES \citep{2007MNRAS.380.1335M,2010NewA...15..460M}, and our code, which is based on the code of \citet{1998ApJ...496L.121K}.  Each of these codes has been developed independently, and can be quite different in their numerical treatment of radiative transfer, subgrid scale models, advection schemes, as well as input physics such as equations of state and opacity.

Simulations have been successfully applied to the outer convection zone in the Sun and in a select group of individual stars.  Such simulations have led to improved p-mode oscillation frequencies in the outer layers of the models where errors in the sound speed contribute to discrepancies between the observed and model frequencies.  Including some turbulent effects has lead to improved frequencies in standard solar models \citep{1999A&A...351..689R,2002ApJ...567.1192L} as well as $\eta$ Bo\"otis and $\alpha$ Centauri \citep{2006ApJ...636.1078S,2007IAUS..239..388S}.

Simulating individual stars has been useful for understanding how convection affects stellar structure in particular cases, and a natural extension of this strategy is to simulate surface convection across the HR diagram.  Some progress has been made in parameterizing the turbulent dynamics in the SAL in this way.  For example, \citet{1995LIACo..32..213L,1998IAUS..185..115L,1999A&A...346..111L} and \citet{1999ASPC..173..225F} have undertaken extensive efforts to map the mixing efficiency (or effective $\alpha$ for MLT) using 2D RHD simulations, and constructing 1D envelope models by matching the specific entropy of the models and simulations.  Recently, \citet{2011ApJ...731...78T} used 3D simulations to map convective efficiency in a range of the HR diagram that includes main-sequence stars and giants.  \citet{1999ASPC..173..233T} have also attempted to extract a mixing length parameter from simulations by matching averaged 3D simulation stratifications to 1D envelope models and modifying the turbulent pressure.  Simulations have also been applied to the formation of spectral lines and interferometric observables. \citet{2000A&A...359..729A,2009ARA&A..47..481A} famously used 3D spectral synthesis to study solar line profiles for abundance analysis, but similar techniques have since been used on the sun \citep[e.g.][]{2010A&A...514A..92C,2011SoPh..268..255C} and other stars \citep[e.g.][]{2009A&A...501.1087R,2007A&A...469..687C,2009MmSAI..80..719C,2009A&A...506.1351C,2010A&A...524A..93C,2009A&A...508.1429W,2010A&A...513A..72B}.

While these applications have been largely successful, the accuracy of simulations is limited by what physical processes are included, and in what manner they are represented numerically.  One way to get a handle on the uncertainty is through comparing different approximations for the same physical process. Encouragingly, comparisons by \citet{2009MmSAI..80..701K}, and more recently \citet{2012A&A...539A.121B}, show that existing codes are in good agreement for simulations of the sun.  There are, however, apparent differences in the thermal structure.  Because there are numerous differences between each of the codes, the source of the differences cannot be traced to a particular method or approximation.  \citet{2004A&A...421..755V} and \citet{2011A&A...535A..22C} have looked at the effect of gray and non-gray opacities in 3D simulations, but their comparisons necessarily used the same radiative transfer method.  

The goal of this work is to examine the extent to which the radiative transfer method affects the resulting thermal structure and gas dynamics of the simulation, while keeping the rest of the physics the same.  In Sections \ref{sec:code} and \ref{sec:rt} we describe our code and the two radiative transfer schemes that we have implemented.  Section \ref{sec:compare} compares results from simulations computed with each of the solvers.

\section{The Radiation Hydrodynamics Code} \label{sec:code}

Stellar convection zones can span many scale heights (the solar convection zone exceeds 20 pressure scale heights) and evolve over a range in timescales.  The thermal relaxation time of an entire convection zone is too large to be simulated entirely with local RHD LES, so the simulation domain is limited to several scale heights near the surface.  Energy transport transitions from fully convective at the bottom of the computational domain, to fully radiative at the top.

The radiation hydrodynamics code is based on the code of \citet{1998ApJ...496L.121K}.  It solves the equations of hydrodynamics as described by \citet{2003MNRAS.340..923R}, but the radiative term {\qrad} is computed using either the 3D Eddington approximation or long-characteristic ray integration.  Radiative transfer is treated with the diffusion approximation in the lower part of the domain ($\tau > 10〗^4$) where it is valid.

Two popular simplifications to the governing equations are the Boussinesq and anelastic approximations.  In both cases, sound waves are filtered out which relaxes the time step restriction imposed by the CFL condition.  Neither of these two approximations are suitable for simulating realistic stellar surface convection. The Boussinesq approximation is not valid because it does not permit strong stratification and the Mach number must be small.  The anelastic approximation, which can be used in a heavily stratified flows, is valid only in the deep convection zone but is unsuitable for convection near the SAL where the Mach number is high.  Instead, we solve the fully compressible Navier-Stokes equations:

\begin{equation} \frac{\partial \rho}{\partial t} = - \nabla \cdot \rho \vec{v} \end{equation}

\begin{equation}\frac{\partial \rho \vec{v}}{\partial t} = - \nabla \cdot \rho \vec{v} \vec{v} - \nabla P + \nabla \cdot \Sigma + \rho \vec{g} \end{equation}

\begin{equation} \frac{dE}{dt} = \nabla \cdot \left[ (E+P)\vec{v} - \vec{v} \cdot \Sigma + f  \right] + \rho \vec{v}\cdot\vec{g} + Q_{\rm rad} \end{equation}

where $E = e + \rho v^2/2$ is the total energy density and $\rho$, $\vec{v}$, $P$, $e$, and $\vec{g}$ are the density, velocity, pressure, specific internal energy, and gravitational acceleration, respectively. The viscous stress tensor for a Newtonian fluid is $\Sigma_{ij} = \mu ( \partial v_i / \partial x_ j + \partial v_j / \partial x_i) - 2/3 \mu( \nabla \cdot \vec{v})\delta_{ij}$.  The radiative term {\qrad} is computed using one of the two radiative transfer schemes described in Section \ref{sec:rt}.  Explicitly coupling the radiative term, {\qrad}, as a source term in the hydrodynamics equations imposes a restriction on the time step.  Switching radiative transfer solvers  changes the method for computing {\qrad}, but does not ease the restriction.

The nature of stellar convection makes it impossible to directly simulate the range in dynamical scales. LES resolve the scales where most of the turbulent energy transport takes place, however, it is necessary to account for the energy transport on unresolved scales. We accomplish this by adjusting the dynamic viscosity to account for sub-grid scale stresses according to \citet{1963MWR....91...99S}.

Where possible, we keep the physics of the simulation consistent with that of the 1D stellar evolution code YREC \citep{2008Ap&SS.316...31D}.  The code uses OPAL equation of state and opacity tables \citep{2002ApJ...576.1064R}, and the \citet{2005ApJ...623..585F} tables for low temperature opacities.  The two simulations presented in this work use Rosseland mean opacity computed from the tables.  The opacity and equation of state are both for the GS98 mixture \citep{1998SSRv...85..161G}.  Further, we treat radiative transfer in local thermodynamic equilibrium (LTE) with gray opacities, but these are not limitations of the methods; both can be extended to account for non-gray effects using methods described by \citet{1982A&A...107....1N}, \citet{1994A&A...284..105L}, and \citet{2004A&A...421..741V}.

\section {Radiative Transfer in 3D Simulations} \label{sec:rt}

There are numerous methods for the treatment of multidimensional radiative transfer (see \citet{2008PhST..133a4012C} and references therein for a review of various methods).  Radiative transfer in 3D was first presented by \citet{1985pssl.proc..215N}.  Today, the most commonly used techniques for 3D radiative transfer are long- \citep{1970ApJ...161..255C,1967ApJ...150L..53A} and short- \citep{1988JQSRT..39...67K} characteristic methods.  Both of these methods share the same strategy for 3D spatial coverage, wherein they combine 1D solutions along a few selected rays to sample all $4 \pi$ steradians.  In the following sections we describe the methods of long-characteristic ray integration and the 3D Eddington approximation, which represent two fundamentally different approaches to achieve spatial coverage. Domain coverage is important because the transfer of radiation is not a local problem in the SAL.

Perhaps the most significant challenge for any 3D radiative transfer solver is to adequately account for the local angular variation of {\qrad} (see Section \ref{sec:raygeom}). The two methods that we compare approximate 3D domain coverage in very different ways. Ray integration techniques estimate {\qrad} by solving the transfer equation along 1D characteristics, but they have potentially poor spatial coverage that is achieved through combining a few 1D solutions. Achieving complete domain coverage in this manner is not computationally feasible, so characteristic methods are necessarily undersampled. The 3D Eddington approximation approaches the problem entirely differently. In it, radiative transfer and domain coverage are approximated at the outset through moment equations that are closed at the first order. In the following sections we investigate the extent to which these two approximations affect the structure of the SAL. The accuracy of the spatial coverage in moment methods such as the 3D Eddington approximation can be enhanced by taking higher order moments, which is analogous to improving the spatial coverage of ray integration methods with additional characteristics.

\subsection{3D Eddington Approximation} \label{sec:3deddington}

As an alternative to characteristic based ray integration methods, the radiative transfer equation can be approximated through moment equations as described by \citet{1966PASJ...18...85U}:

\begin{equation} \label{eqn:edd}
\nabla \cdot \left( \frac{1}{3 \rho \kappa} \nabla J \right) - \rho \kappa J + \rho \kappa B = 0.
\end{equation}
and the net radiative heating or cooling is:
\begin{equation}
Q_{\rm rad} =  4 \rho \kappa (J-B)
\end{equation}

The 3D Eddington approximation has been used in codes applied to other astrophysical regimes \citep[e.g.][]{1980ApJ...236..619B,1989CeMec..45...85B}, although to our knowledge our code is the only application of it to simulations of near-surface convection.  The Eddington approximation describes isotropic radiation exactly in both optically thin and  thick regions, and is not equivalent to the diffusion approximation in optically thin regions \citep{2003rtsa.book.....R}. Further, although we use the Planck function as the source function, the 3D Eddington approximation does not strictly require LTE.

The 3D Eddington approximation has advantages over other methods, afforded by its simplicity and speed.  One of the advantages of the 3D Eddington approximation is that the nature of its formulation alleviates concern that a nearby source or sink of energy will be missed as a result of limited domain coverage.  This is not to say that the spatial domain is fully sampled (it is still defined by first order moments), but it is complete in that it includes information from all other points and not just those that would be sampled by characteristics.  The angular resolution is limited by the first order moments, but it is sensitive to the complete domain because of the iterative nature of the numerical solvers.  At a particular location at an instant in time, the local {\qrad} computed from ray integration only includes information from along the characteristics, which could potentially miss an important source or sink.  Two grid points that are not connected by a ray can only influence each other indirectly.  In simulations of convection in the SAL, this risk is largely mitigated by the time evolution, or by rotating the undersampled bundle of rays in a manner similar to \citet{2003ASPC..288..519S}.  Complete domain sensitivity can be ensured by iterating on the ray solution, but this adds to the already high computational cost. The 3D Eddington approximation achieves domain coverage without requiring such methods.
 
The second advantage is that the 3D Eddington approximation can be solved with a simple and fast numerical method.  This is in a large part because the solution is computed directly on the hydrodynamic grid, and there is no need to define additional grids for radiative transfer along each characteristic.  Aside from the computational convenience of having the radiation and hydrodynamic solvers share the same grid, errors introduced from interpolation are also avoided.  Interpolation is likely not the dominant source of error, and is almost certainly smaller than the error introduced by the 3D Eddington approximation for spatial coverage or the undersampled numerical quadrature, so avoiding interpolation is primarily a performance advantage.  Interpolations in ray integrations can be avoided altogether by casting rays so that they always intersect the hydrodynamic grid, but flexibility in ray placement would be lost, resulting in reduced domain sensitivity.

\subsubsection{Iterative Solvers}

Solving large sets of partial differential equations has been dealt with in great detail, and there exists a rich body of work from which we can develop solvers for the anisotropic inhomogeneous Helmholtz equation (see \citet{2003ccdouglas.book.....D} and references therein).  The 3D Eddington approximation, in conjunction with assuming LTE to set the source function, results in a linear relationship for the mean intensity, $J$. In principle, Equation (\ref{eqn:edd}) can be written as a large linear system and solved directly, but this is not done in practice because of the enormous computational cost.  Much faster iterative (relaxation) methods are used instead.

While classical relaxation techniques are sufficient to solve Equation (\ref{eqn:edd}), methods with better convergence properties, such as alternating-direction-implicit (ADI) and multigrid methods, reduce the computational burden significantly.  Multigrid methods are particularly powerful, as they can render the computational cost proportional to the number of grid points \citep{1992nrfa.book.....P}.  In the following sections we describe two solvers that we use in the code.  Note that the two methods are exactly equivalent in that they produce solutions to Equation (\ref{eqn:edd}) that agree to within the specified error threshold.  Because the solvers result in the same solution, selecting a solver is based on numerical stability and parallel performance considerations.  

First introduced by \citet{1955JAICHE....1...505S} and \citet{1955JSIAM....3...28S}, ADI methods are useful on regularly spaced multidimensional grids.  Both \citet{1992nrfa.book.....P} and \citet{2003ccdouglas.book.....D} describe the method in detail.  Each ADI iteration is comprised of three one-dimensional implicit problems, obtained by splitting the terms in equation \ref{eqn:edd} according to spatial dimension.  For each spatial dimension, the derivatives are treated implicitly only for that dimension.  The terms corresponding to each dimension represent a series of tridiagonal linear systems, for which there are many fast solvers \citep[e.g.][]{laug}.

The systems are solved sequentially in `sweeps' along each dimension. Completing the three directional sweeps provides an improved estimate of the solution.  The final solution is achieved by iterating until the desired error threshold is reached.  The convergence properties of the ADI technique are sensitive to the initial guess, asymmetry of the flow, and the problem size.  Our tests show that convergence is not always linear with problem size, which can become cost-prohibitive for very large grids. 

Unlike the ray integration scheme, the ADI solver does not present a natural parallelism that is as easy to exploit.  Although each tridiagonal system within a directional sweep is independent, the successive iterations can result in very high communication costs.  To achieve scalability on distributed memory machines, we use a process-scheduling algorithm with asynchronous communication to overlap computation and communication. While it is not our intent to discuss parallel algorithms in detail, we point out that \citet{Povitsky:2002:PAS} and \citet{2003ccdouglas.book.....D} discuss strategies for parallel ADI algorithms.

We have also implemented a so-called \textit{optimal algorithm} using a geometric multigrid method.  Such methods solve partial differential equations using a recursive procedure applied to a hierarchy of grids. Multigrid solvers achieve better convergence rates by reducing low-frequency errors on coarser grids and interpolating the solution to finer grids \citep{2003ccdouglas.book.....D}. The solution is mapped between grids using smoothing, restriction, and interpolation operators.  The two major advantages of this method are stability and scalability. 

The multigrid method is much more stable than ADI in regions of anisotropic flow.  The ADI method requires a good initial guess for the solution.  Using the solution at the previous time step for the initial guess is usually is sufficient, but if the flow is highly anisotropic, the ADI method can fail to converge.  Unlike the ADI solver, the multigrid solver does not require a good initial guess for the solution.  This makes the solver very stable, even for highly turbulent flows, such as granulation in giants and early type stars.  

The scalability of the solver is \textit{optimal} in the sense that in scales linearly with problem size.  That is, is solves the equation with operational cost proportional to the number of unknowns.  Our tests confirm that the solution is achieved with $\mathcal O (N)$ operations without a good initial guess. Although not strictly required, a good initial guess improves the speed of the solver.

Our multigrid method is based on the `full multigrid' solver described by \citet{1992nrfa.book.....P}, modified to solve the anisotropic inhomogeneous Helmholtz equation in 3D, with periodic boundary conditions on the horizontal walls and Neumann boundary conditions at the top.  It employs W-cycle structure for multigrid cycles, and Gauss-Seidel smoothing with red-black ordering. As yet, our multigrid solver is not parallelized for distributed memory, which limits its use to shared memory machines, however, a full parallel version of this solver can be implemented following the strategy outlined by  \citet{2003ccdouglas.book.....D}.

\subsection{Long Characteristic Ray Integration} \label{sec:rayint}

Ray integration methods are attractive because they solve the integro-differential radiative transfer equation along rays (or characteristics), which provides a robust solution along a 1D path.  Extending the solution to 3D, however, requires solving the transfer equation along many ray paths that cover every direction.  Good domain coverage can be achieved when many rays are used, but the computational cost of this is typically too high.  To alleviate the computational burden, only a few rays are used.  

In contrast to the simplicity of the 3D Eddington approximation, long-characteristic ray integration is comprised of a ray integration step and a spatial integration step that must be carried out at each location in the computational domain.  The first part of the solution is to perform the 1D radiative transfer calculations at various angles through the domain.  The second part of the solution is to integrate the result from the 1D rays at each grid point over the unit sphere with numerical quadrature.  The details of how these two steps are implemented can have an effect on the final solution.  We discuss our implementation in Sections \ref{sec:rayrad} and \ref{sec:raygeom} below.

In terms of computational performance, our 3D Eddington solvers outperform the ray integrators (see Section \ref{sec:performance} for a detailed comparison), but ray integration presents a natural parallelism that can be exploited. In modern computing, scalability is often more important than the inherent cost of the algorithm.  \citet{2006A&A...448..731H} and \citet{2010NewA...15..460M} describe parallel algorithms for ray integration that can be applied to distributed memory machines using domain decomposition.  The good performance over a large number of distributed memory compute nodes makes ray integration an attractive option for modern codes.

\subsubsection{Solving the Transfer Equation} \label{sec:rayrad}

Following the description of \citet{2003ASPC..288..519S}, we adopt the \citet{Feautrier1964} representation of the transfer equation for the numerical solution:

\begin{equation} \label{eqn:feautrier}
\frac{d^2P}{d\tau^2} = P - S,
\end{equation}

where $S$ is the source function (taken to be the Planck function), $P = 1/2[I_+(\tau)-I_-(\tau)]$ and $I_+(\tau)$ and $I_-(\tau)$ are the outgoing and incoming specific intensities at optical depth $\tau$. The local heating/cooling {\qrad} is computed by integrating in each direction:

\begin{equation} \label{eqn:qplusminus}
Q^+ = I^+ - S, \hspace{4mm}
Q^- = I^- - S,
\end{equation}
and the contribution to the net radiative heating or cooling for the ray is
\begin{equation}
Q_{\rm rad} = Q^+ + Q^-.
\end{equation}

We use the Hermitian differencing method of \citet{1976JQSRT..16..931A}, which achieves 4th order accuracy while maintaining essentially the same cost as a 2nd order scheme.  We found similar results for the final {\qrad} when using the 2nd and 4th order schemes, but the sensitivity on the result may depend on the spatial resolution of the simulation, as well as the method for interpolating between the hydrodynamic and radiation grids (discussed below).

We define all ray characteristics to begin at the top of the box and integrate into higher optical depth. Rays are cast in directions defined by two angles $(\theta,\phi)$.  The $\theta$ angle is measured with respect to the vertical direction, and $\phi$ is the azimuth angle. Each ray direction corresponds to a bundle of parallel rays that traverse the computational domain. In order to compute the formal solution along a given characteristic, the source function and optical depth must be known at each point along the ray.  From this, the radiative heating or cooling {\qrad} is computed at each of the points where the source function and optical depth are known. Because the ray integration is performed along lines that are not usually coincident with the hydrodynamic grid (the vertical ray is an exception), the required thermodynamic variables must be mapped from the hydrodynamic grid to the radiation grid.

\begin{figure}
\epsscale{1.0}
\plotone{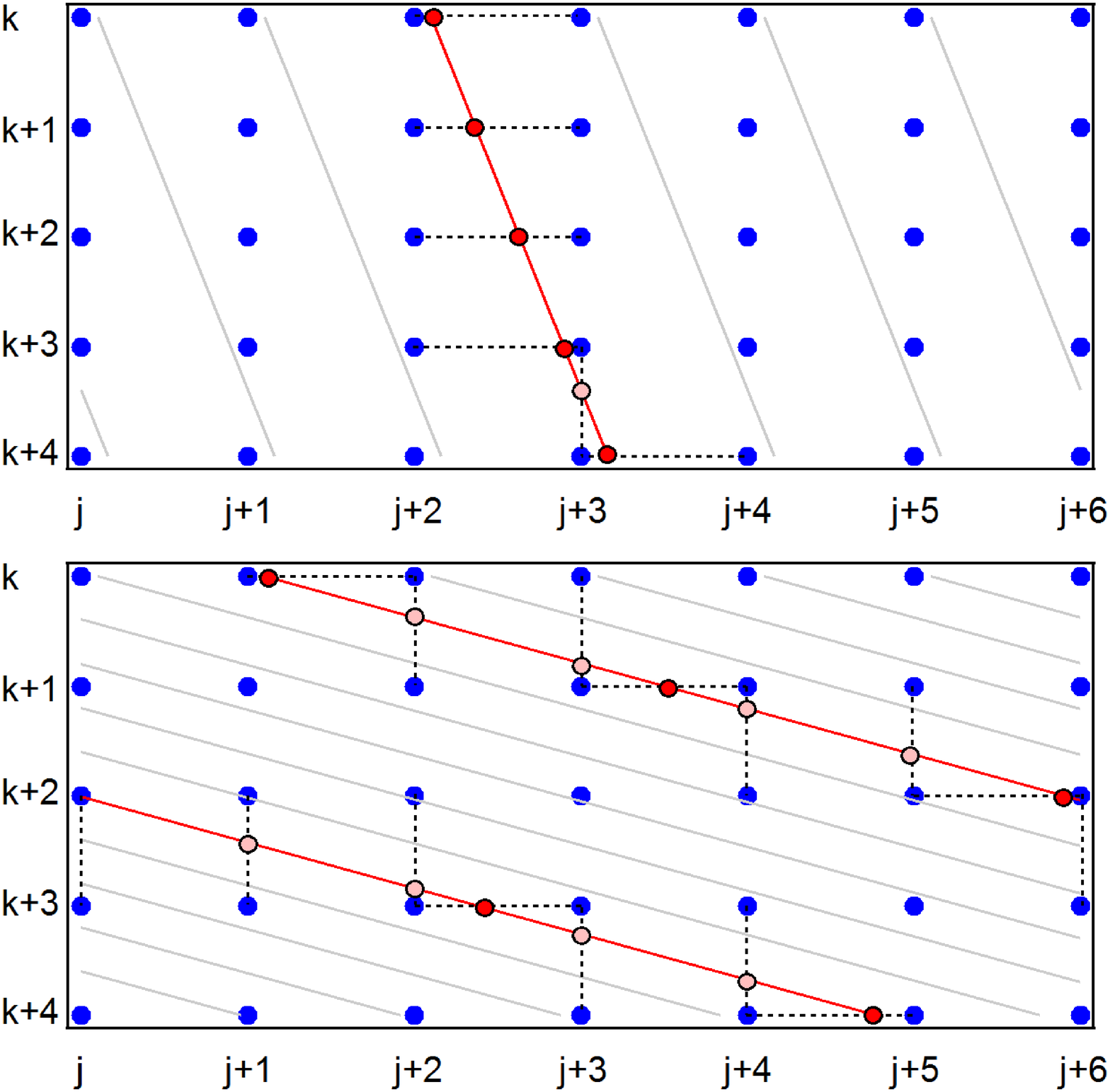}
\caption{A 2D schematic illustration of a steep (top) and shallow (bottom) long ray cast through a periodic domain.  Mapping between the ray and hydrodynamic grids requires interpolating data from the hydrodynamic grid (blue points) to the red points.  The solution along a characteristic (red points) is only required at intersections of horizontal planes (solid red points) but all intersection points are used to avoid very large steps in $\tau$ for shallow rays.}
\label{fig:gridmapping}
\end{figure}

The hydrodynamic grid is defined by the regular 3D Cartesian mesh that specifies the locations of zone-centered quantities.  The temperature, density, and opacity are tracked on this grid.  The radiation grid is defined by the intersection points of a parallel ray bundle with the hydrodynamic grid.  Intersection points occur in 2D planes between the zone centers of the hydrodynamic grid, so a 2D interpolation is required at each intersection to determine the required thermodynamic quantities for ray integration.  Figure \ref{fig:gridmapping} shows a schematic with one ray intersecting the hydrodynamic grid at a steep and shallow angle.

A ray always has a vertical component (we do not use periodic rays), so the final solution only requires points in the horizontal planes of the hydrodynamic grid.  Very shallow rays can traverse long distances between successive intersections of the horizontal planes of the hydrodynamic grid, so we include intersections of the vertical planes as well to ensure that each ray is adequately sampled before performing the Feautrier integration.  The intersections with vertical planes are only included if the distance to the next horizontal plane is large.  This restriction avoids having points that are spatially very close together. The result is that the forward mapping (from the hydrodynamic to radiation grid) includes most intersections of the ray with the hydrodynamic grid, but the reverse mapping (from the radiation to hydrodynamic grid) only includes intersections of the ray with horizontal planes of the hydrodynamic grid.

The mapping between the two grids introduces a smoothing to the solution twice.  First, the hydrodynamic quantities (in this case $\rho \kappa$ and $T$) must be interpolated to points along the ray bundles.  The second smoothing occurs with the reverse transformation of interpolating the computed {\qrad} back to the hydrodynamic grid.  The smoothing effect can be somewhat mitigated by employing higher order interpolation schemes, such as bicubic or distance weighted parabolas, but this will add to the computational expense.

\subsubsection{Spatial Coverage} \label{sec:raygeom}

After the ray integration step, the contribution to the radiative heating or cooling is known for each ray direction, and the result of all rays at each grid point must be integrated over the unit sphere. To completely cover the 3D domain, rays need to be cast along every line of sight over $4\pi$ steradians, at every point in the grid. This is far too computationally demanding, so we use only a few rays which are integrated over $4\pi$ steradians using a quadrature rule.  There are numerous quadrature rules available (for example, see \citet{Abramowitz/Stegun/1972} or \citet{Antia1991}) and the choice of rule may affect the total radiative flux in the simulation.  

For local RHD simulations we can take advantage of azimuthal symmetry by spacing rays at fixed intervals of $\phi$, and the integration reduces to a 1D quadrature along the azimuth, and a second 1D quadrature along the cosine-angle. If the rays equally spaced in azimuth, the weights of rays with the same $\theta$ are equal, and the unit sphere integration is reduced to a one dimensional calculation. Using a quadrature rule, the integration is a weighted sum of the radiative heating at each grid point.  Adding to the number of rays improves the accuracy of the unit sphere integration, and achieving good spatial coverage is important for accurately calculating the local and total energy flux.  

Typically only a few rays are used to keep the computational expense manageable.  For example, the STAGGER code uses 5 rays which are rotated by 15 degrees every time step \citep{2003ASPC..288..519S}, or 9 rays in recent work \citep{2012A&A...539A.121B}. The solution is quite sensitive to integration over $\theta$ because the vertical energy flux is the largest contributor to the total flux.  If the ray directions are not constrained by the hydrodynamic grid, the theta angle of the rays can be chosen to correspond to a desired quadrature rule.  In this work we use Gauss-Radau quadrature \citep{Hildebrand-1956} with one vertical ray and four rays inclined at $\cos \theta = 1/3$, which is the same as that used in the STAGGER code (in \citet{2003ASPC..288..519S}).  The vertical ray has a weight of $1/4$ and the four inclined rays have a combined weight of $3/4$.

\subsection{Variation of Local Radiative Heating and Cooling} \label{sec:variation}

Spatial coverage is critical for radiative transfer in simulations of near-surface convection.  A radiative transfer solver must have good spatial coverage because of the large asymmetries that convection introduces in the flow, and because the SAL spans the transition from optically thick to thin where radiative transfer becomes non-local.  The cost of good spatial coverage can be problematic for 3D RHD simulations, so approximations are made to the angular resolution to keep the computational costs affordable. The 3D Eddington approximation simplifies spatial coverage through moments, while they ray integration relies on numerical quadrature.  In both cases the result can be inaccurate if the function is highly variable or asymmetric.  Both of these properties exist for the radiative heating in the SAL and pose a challenge for any affordable 3D radiative transfer method.

The stellar flux is predominantly vertical (introducing large asymmetry in the function) and large horizontal variation can occur from the high contrast of hot upflowing granules and cool inter-granular lanes.  The way that radiative transfer schemes deal with this is a large source of uncertainty and limits the absolute accuracy of the methods. For example, spatial quadrature rules based on different sets of polynomials will assign different weights to the vertical and horizontal rays, which can result in different estimates for the total energy fluxes from the same opacity and source function.  This is a significant source of uncertainty, and one that can only be eliminated by using a very large number of rays.

\begin{figure*}
\begin{minipage}[b]{0.3\linewidth}
\plotone{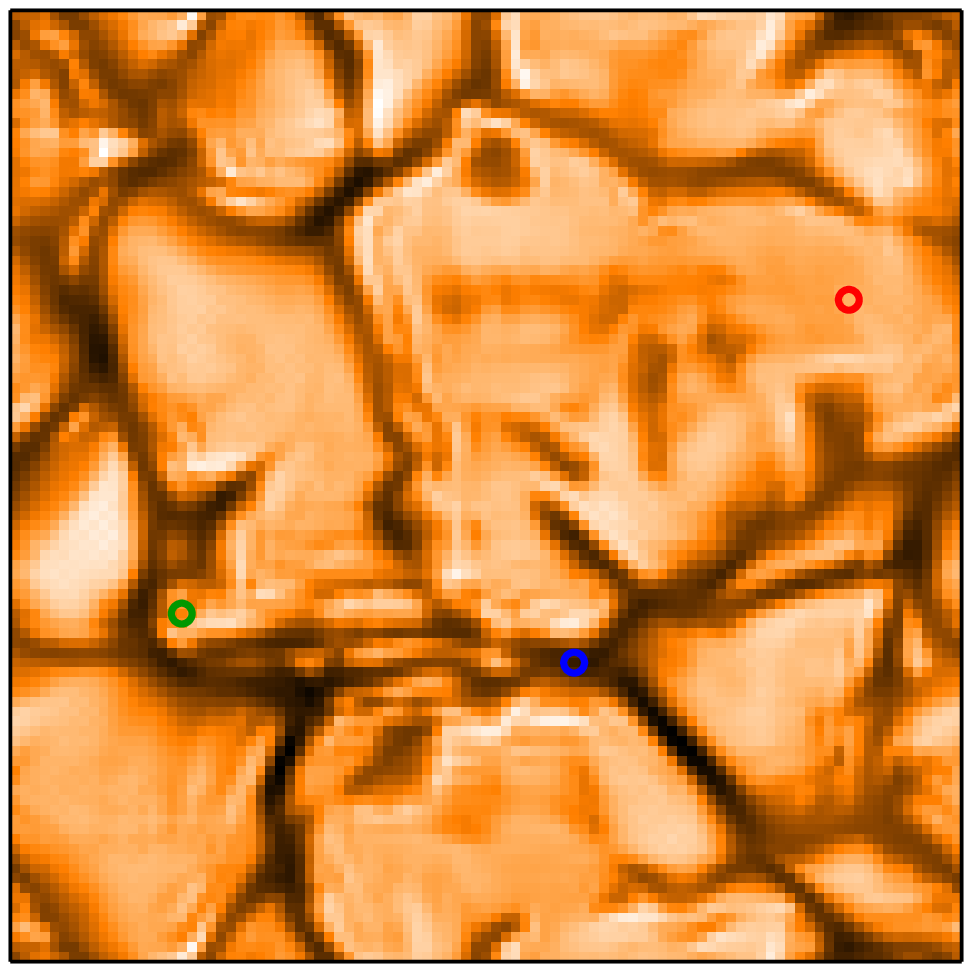}
\end{minipage}
\hspace{1mm}
\begin{minipage}[b]{0.3\linewidth}
\epsscale{1.09}
\plotone{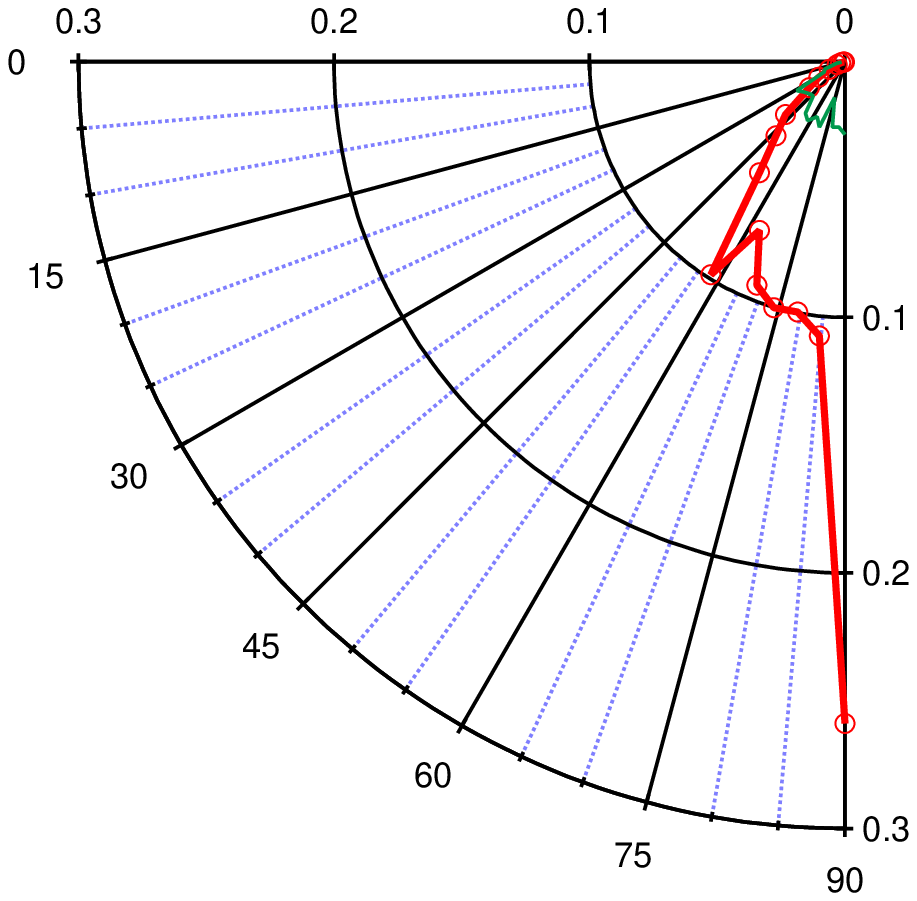}
\end{minipage}
\hspace{1mm}
\begin{minipage}[b]{0.3\linewidth}
\epsscale{1.09}
\plotone{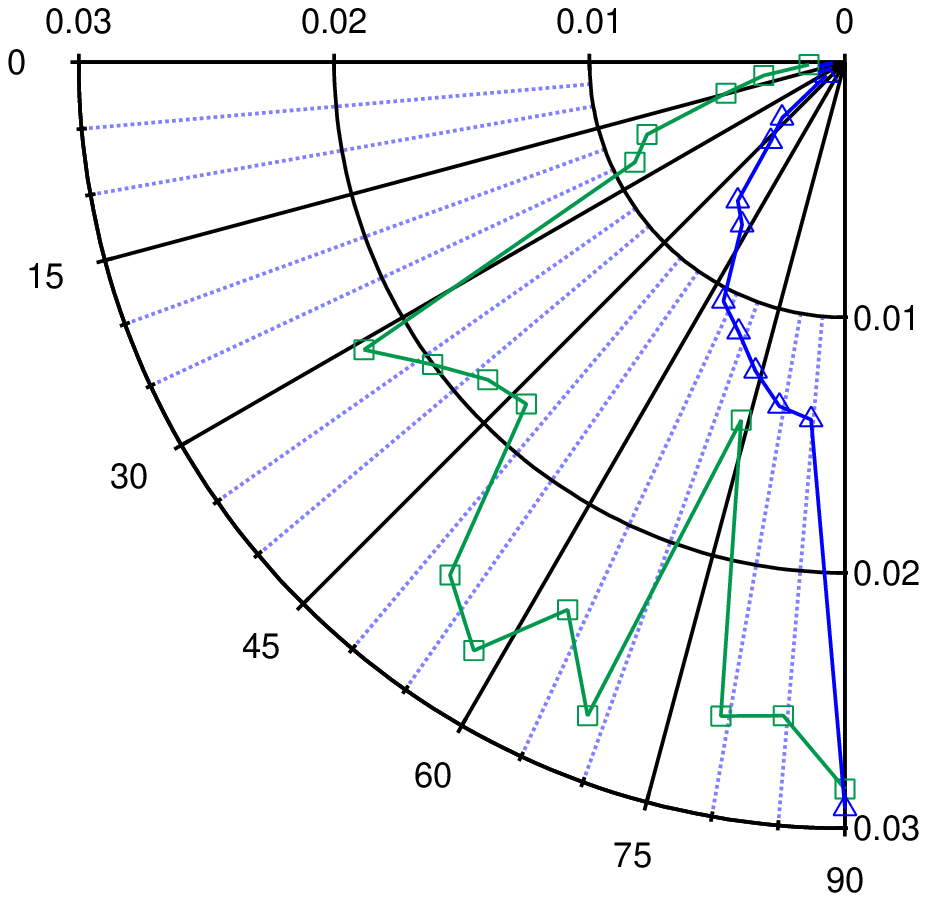}
\put(-80,150){\rotatebox{0}{\scriptsize$Q_{\rm rad}$}}
\put(-233,150){\rotatebox{0}{\scriptsize$Q_{\rm rad}$}}
\put(-3,88){\rotatebox{270}{\scriptsize$Q_{\rm rad}$}}
\put(-155,88){\rotatebox{270}{\scriptsize$Q_{\rm rad}$}}
\end{minipage}
\caption{The azimuth-averaged angular variation in the local radiative heating and cooling {\qrad} (in dimensionless units normalized to the stellar flux) for three locations (marked in the left panel) near the optical surface.  The area enclosed by each {\qrad} curve is the local net heating or cooling. The left panel shows the locations (large upflow in red, small upflow in green, and downflow in blue) for the local {\qrad} profiles in the other two panels.  The {\qrad} profiles can exhibit large asymmetry and variation over small changes in angle.  Using quadrature rules with two or three angles can systematically under- or over-estimate the stellar flux.}
\label{fig:localqrad}
\end{figure*}

To show how the ray integration method is sensitive to spatial coverage, we compute {\qrad} using the long-characteristic ray solver with $65$ rays, on a snapshot from the thermally relaxed 3D Eddington simulation.  Note that the number of rays is far in excess of what is typical or affordable for a simulation.   In the simulations, the result of the ray calculation is integrated over the unit sphere to achieve the net {\qrad}, which is then applied to the Navier Stokes equations.  In this case, however, we average over azimuth and present the local {\qrad} as a function of cosine-angle.  Figure \ref{fig:localqrad} presents several examples of the local {\qrad} corresponding to three different regions of surface granulation at the $\tau=1$ surface.  We present the local {\qrad}  in dimensionless units, normalized to the total stellar flux. Even with 65 rays, the variation in local {\qrad} is not always well resolved.

The curve in the center panel of Figure \ref{fig:localqrad} shows {\qrad} for the center of a granule, where the local flux is dominated by the vertical ray.  This is expected, as it is the hot upflows that contribute the majority of the stellar flux.  Because the center of a granule is hot and surrounded by hot gas, the heating from horizontal directions is quite small relative to the vertical flux.  In upflowing regions such as this, the large variation of the local {\qrad} function can make it difficult to determine the local energy flux if only a few rays are used.

In the narrow intergranular lanes, the gas is cool and surrounded by much hotter upflows.  Here, the local energy flux is still dominated by the vertical direction, but the magnitude of the flux is much smaller than in the hot upflow.  Although there is still a net heat flux vertically, nearby upflows are heating the gas from shallower (horizontal) angles.  The intergranular lanes show a large asymmetry of the local {\qrad} function, and the local flux can change sign at small angles.  Both of these properties can be troublesome for a sparsely sampled quadrature rule.

Regions of mixed flow, such as a small upflow that is near to an intergranular lane, show characteristics of both the hot and cold extremes described previously.  The mixed flow example presented in Figure \ref{fig:localqrad} shows a similar vertical flux as in the downflow region, but has significant heating from angles far from vertical.  Furthermore, the variation in heating as a function of angle varies considerably over small angular increments.  The local flux is much smaller than that from the center of a large granule, but the local {\qrad} function is still very asymmetric and dominated by the vertical direction.

The three examples in Figure \ref{fig:localqrad} show the angular variation in {\qrad} at an instant in time, and do not represent the bulk flow.  Because of the highly variable nature of the local {\qrad} profile, placing rays to adequately sample the vertical flux may be important for accurately determining the radiative energy flux.

\subsection{Performance Considerations} \label{sec:performance}

Our comparison in Section \ref{sec:compare} of the moment-based 3D Eddington approximation with the under-sampled ray integration method shows that they result in largely equivalent stratifications through the stellar SAL.  Current 3D RHD simulations are limited by computational power, so it is prudent to compare the costs of these two methods that produce equivalent results.  Numerical solvers for moment based methods are fast, and offer significant computational savings.

\begin{figure}
\epsscale{1.0}
\plotone{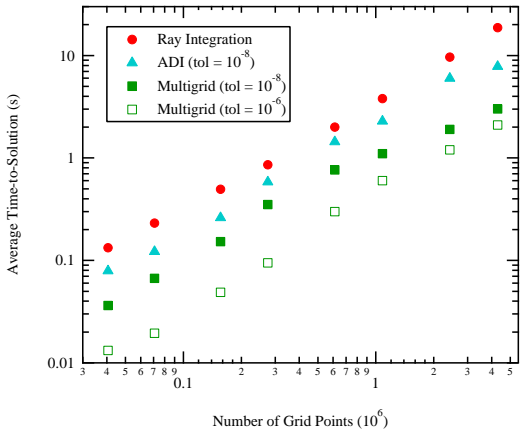}
\caption{Performance comparison of the ray integrator with the ADI and multigrid solvers for the 3D Eddington equation.  All tests were performed on the same hardware using a single processor. The 3D Eddington solvers have a clear performance advantage over ray integration.}
\label{fig:performance}
\end{figure}

In our experience, the 3D Eddington solvers typically out-perform the ray integrators by a significant margin.  Figure \ref{fig:performance} shows the time-to-solution for simulation domains of varying size.  The radiation solvers are applied only to the top of the simulation domain where the temperature gradient is significantly superadiabatic.  The cases presented here correspond to a simulations with a superadiabatic domains that are 65 zones deep and with varying horizontal sizes.  All simulation runs for the comparison were performed on the same hardware using a single processor.

The ray integration solver presented in figure \ref{fig:performance} is our most economical one, which uses 5 rays with bilinear interpolation.  Note that the cost of additional rays is not simply proportional to the number of rays because shallow rays traverse a longer path through the simulation domain.  A significant improvement in performance could be realized by adopting the scheme described by \citet{2012JCoPh.231..919F}, where only ray intersections with horizontal planes are included, and by aligning the azimuthal angle of the rays with the hydrodynamic grid, thereby reducing interpolations to one dimension. The smaller computational cost of the one dimensional interpolations comes with the consequence of reduced domain sensitivity because the azimuth angle of the rays could no longer be rotated each time step.  Because the ray integrator is not iterative, for a given ray configuration the amount of computational work is linearly proportional to the size of the grid.  Increasing the horizontal extent of the simulation domain results in the expected linear increase in computational cost.

Similarly, the cost of both the ADI and multigrid solvers for the 3D Eddington equation show a roughly linear proportionality with grid size.  A good initial guess is necessary for the ADI method to achieve linear scaling, but no such guess is necessary for the multigrid method.  The iterative 3D Eddington solvers have a tolerance parameter that defines how close the numerical solution matches equation (\ref{eqn:edd})  We see a performance benefit by adjusting the tolerance parameter from $10^{-8}$ to $10^{-6}$.

The speed difference between the moment and characteristics methods can make a tangible difference.  Given that the computational cost of a 3D radiation hydrodynamics simulation is typically dominated by the radiation solver, the transport of radiation is simplified, for example, by limiting the domain coverage, or by limiting the frequency coverage to a few opacity bins. The computational cost savings afforded by moment based methods, which appear to offer similar domain coverage as under-sampled ray integration, could be used to better cover the frequency domain using opacity binning or sampling techniques.

\section{Comparing Radiative Transfer Schemes in 3D Simulations} \label{sec:compare}

We have implemented the two methods described in Section \ref{sec:rt} for computing radiative heating and cooling in a 3D LES.  While differences in the methods are clear, the only way to determine their effect on the thermal structure and gas dynamics from solving the Navier-Stokes equations is to use them in simulations.  Because {\qrad} is essential in determining the thermal structure, we cannot compare methods on the same simulation snapshot at an instant in time.  Instead, we compare statistics gathered from long runs of simulations computed with each radiative transfer solver. 

In this section we compare results of two simulations: one computed with the 3D Eddington approximation using the ADI solver (described in Section \ref{sec:3deddington}), and the other with long-characteristic ray integration (described in Section \ref{sec:rayint}).  The two simulations have the same composition ($Z=0.0169$, $X=0.7366$), energy flux ({\teff}$ \approx 5725 \mbox{K}$), and surface gravity ({\logg} $= 4.438$).  Both simulations span $3.95 \times 3.95 \times 2.85 \mbox{ Mm}^3$ with a resolution of $100 \times 100 \times 215$.  We started each simulation from the same state and evolved them until they achieved a new thermal relaxation based on their respective radiative models.   Statistics were gathered after thermal relaxation.  

The following sections compare various space- and time-averaged quantities from the two simulations.  Figures show the mean values with standard error as a function of height in the simulation domain.  The zero point is defined as the location where the average temperature is equal to the effective temperature.  We use the following definitions for statistical quantities.

\begin{equation}
\bar{q}(z) = \frac{1}{T}\frac{1}{L_x L_y} \int_{t_0}^{t_0+T} \int_{0}^{L_x} \int_{0}^{L_y} {q} \hspace{1mm} {dy} \hspace{1mm} {dx} \hspace{1mm} {dt} ,
\end{equation}
where the quantity has been averaged over a time $T$ in a box of horizontal cross-sectional area $L_x \times L_y$, and $t_0$ is a time after the simulation has relaxed.  The time average is over a sufficient interval to obtain statistical convergence.  The RMS of the same quantity is:
\begin{equation}
q'' = (\overline{q^2}) - (\overline{q})^2.
\end{equation}

\subsection{Radiative Efficiency}

The purpose of the radiative transfer method is to transfer energy within the simulation domain, and determine how energy is ultimately removed (the stellar flux).  The local contribution to radiative heating or cooling comes from applying the radiative transfer solver to the inhomogeneous medium.  Different radiative transfer methods can remove the same total energy from a given stratification, but the energy may be removed in different ways.  In this case, the effective temperature (measured as $F \propto \sigma T_{\rm eff}^4$) remains unchanged, but the thermal structure is different.  This effect would appear as a different transition from convective to radiative energy transport and would affect the shape of the superadiabatic gradient ($\nabla - \nabla_{\rm ad}$).  We refer to this as the ‘radiative efficiency’.  The radiative efficiency determines the rate of transition from convective to radiative energy transport.  A more ‘efficient’ method will start to be super-adiabatic deeper in the star, and will have a higher peak in the superadiabatic gradient.

\begin{figure}
\epsscale{1.0}
\plotone{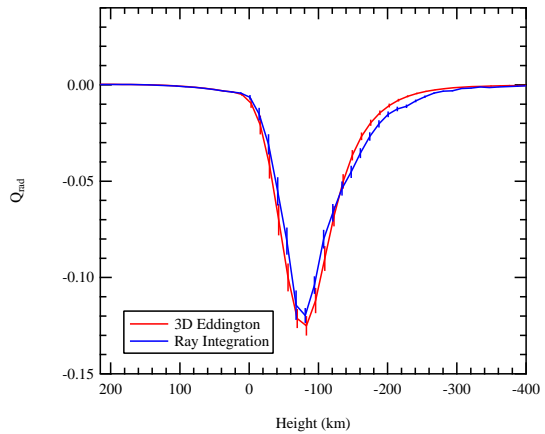}
\caption{Radiative cooling as a function of simulation depth.  The total radiative fluxes (the integral over height) are the same but the shapes of the curves are different.  Lines show the space- and time-averaged mean, with bars showing the standard error.}
\label{fig:qrad}
\end{figure}

Figure \ref{fig:qrad} compares the radiative heating averaged over space (horizontal) and time.  The integral of this function over height is the total energy flux.  Both methods have essentially the same energy flux (their effective temperatures differ by only 30 K, or 0.5\%).    The radiative transfer method has no effect on the adiabatic structure where the diffusion approximation is valid, so the 3D Eddington and ray integration methods have the same deep stratification.  Convection carries the energy flux in the optically thick region, and is not directly affected by the differences in radiative transfer schemes.

\begin{figure}
\epsscale{1.0}
\plotone{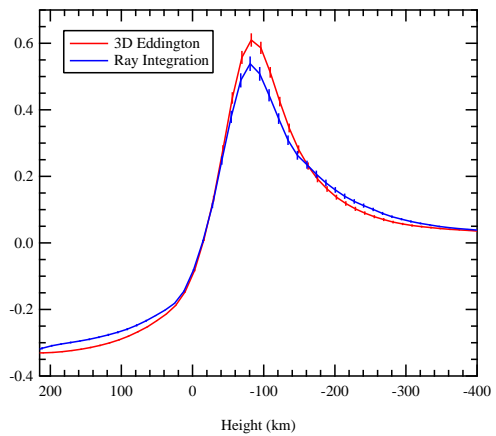}
\put(-465,180){\rotatebox{90}{\large$\nabla - \nabla_{\rm ad}$}}
\caption{Superadiabatic gradient as a function of simulation depth.  The radiative transfer method affects the detailed shape of the superadiabatic gradients.}
\label{fig:sal}
\end{figure}

The superadiabatic gradient in Figure \ref{fig:sal} shows a corresponding change.  A rising parcel of hot gas becomes superadiabatic at greater depth below the SAL in the ray method compared with 3D Eddington.  At the peak of the SAL, however,  the 3D Eddington method radiates more efficiently resulting in a higher peak.  This means that below the SAL, the ray method is more efficient at transferring energy radiatively, but becomes slightly less efficient throughout and above the SAL peak.  The differences are subtle but induce changes to some flow properties  (Figure \ref{fig:wrms}) and the mean stratification (Figure \ref{fig:rho}) in the superadiabatic part of the domain.   The superadiabatic gradient of the 3D Eddington simulation is more negative throughout the atmosphere.

\subsection{Mean Stratification}

\begin{figure}
\epsscale{1.0}
\plotone{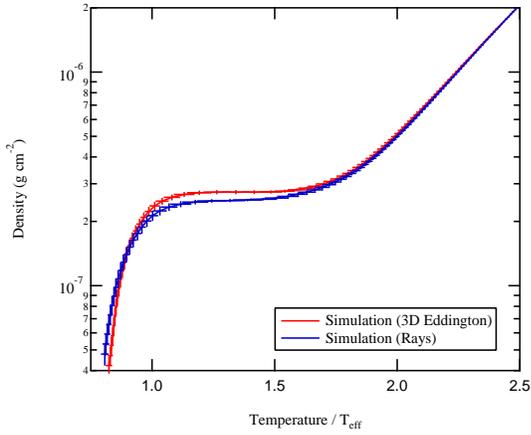}
\caption{The thermal structure (mean density vs temperature) for simulations computed using the 3D Eddington approximation and the ray integration method.  The structures are the same in the adiabatic region but differ through the SAL and atmosphere.}
\label{fig:rho}
\end{figure}

Figure \ref{fig:rho} shows the density stratification as a function of temperature.  Both simulations have the same stratification in the convective region and differ only in the SAL and optically thin atmosphere.  The stratifications agree in the deep region because the two radiative transfer methods compute the same total radiative flux for a given stratification.  The energy transfer in the deep region is entirely convective and is indirectly affected by the radiative energy transport closer to the surface.  The details of the radiative energy transport schemes are different in each simulation, so the shallow region near the top has relaxed to a different state.  The result is that the temperature and density structures of the simulations are distinct.

The density and temperature profiles diverge as the temperature gradient becomes significantly superadiabatic.  The densities differ by a factor of approximately $1.2$ within the SAL (at $T/T_{\rm eff} \approx 1.3$), and by a factor of approximately $1.5$ in the atmosphere (at $T/T_{\rm eff} \approx 0.84$).  Note that this discrepancy is similar to the range exhibited in the comparison of solar simulations by \citet{2009MmSAI..80..701K}.  The differences are likely caused by the different radiative efficiencies.  While the two simulations have the same total energy flux (the integral of {\qrad} over height), their cooling profiles in Figure \ref{fig:qrad} distinct.

Simulations contain many numerical and physical aspects that can affect the results.  The thermal structures for solar simulations computed using different codes with a variety of physical inputs and parameters are indeed different.  The stratifications in Figure \ref{fig:rho} are within the range of what is seen when comparing results from different codes [for example see Figure 1 from \citet{2009MmSAI..80..701K}]. Except for the radiative transfer, all parameters that define the two simulations are the same, so the change in stratification seen here is due entirely to the different radiative transfer schemes.  Both radiative transfer methods result in SAL structures that are realistic.  Although we cannot determine which is more accurate, these results provide a measure of the accuracy of 3D radiative transfer schemes in simulations.

The thermal structure of the atmosphere is often presented as the variation of temperature with optical depth.  In stellar models, this so-called {\ttau} relation is imposed as a boundary condition.  Typically, the Eddington {\ttau} relation is used for computing stellar models, i.e.,

\begin{equation}\label{eqn:eddingtonapproximation}
T^4 = \frac{3}{4} T_{\rm eff}^4 \left( \tau + \frac{2}{3} \right),
\end{equation}

which is purely radiative, and does not include the effects of convective overshoot.  This is one of the reasons that stellar models are inaccurate near the surface. Taking {\ttau} relations from simulations and using them in stellar models is a simple way to include overshoot.

\begin{figure}
\epsscale{1.0}
\plotone{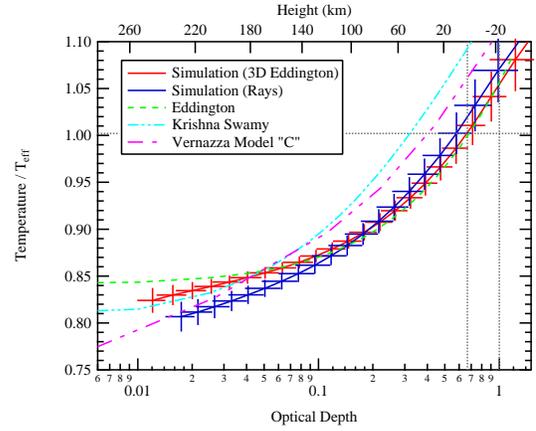}
\caption{{\ttau} relations for simulations computed with different radiative transfer methods.  The atmospheric structures are very similar near $\tau=1$, but show significant differences at low $\tau$.}
\label{fig:ttau}
\end{figure}

We find that the simulated {\ttau} relation is sensitive to the radiative transfer scheme.  While similar to Figure \ref{fig:rho} which shows structural differences in the SAL, Figure \ref{fig:ttau} highlights the differences at small $\tau$.  Several analytic and semi-empirical {\ttau} relations are included in Figure \ref{fig:ttau} for reference.  The simulations deviate from the purely radiative Eddington {\ttau} relation because they include the effects of convective overshoot and turbulence.  These differences are expected for realistic atmospheres, as demonstrated by the semi-empirical \citet{1966ApJ...145..174K} and \citet{1981ApJS...45..635V} models.  Simulations computed by \citet{1997ASSL..225...73T} also show changes in {\ttau} relations as a result of convection.

One of the consequences of convective overshoot is a shift in the location of the mean {\teff} surface with respect to mean optical depth.  The {\teff} surface from the 3D Eddington simulation is very close to $\tau = 2/3$, while the ray integration simulation is at slightly smaller ($\tau \approx 0.6$) optical depth.  The slight differences in the turbulent properties of the simulations (which is a result of the different entropy profiles from using different radiative transfer models) produces slightly different positions of the {\teff} surface.  A shift in the {\teff} surface is not entirely surprising considering that semi-empirical {\ttau} relations have mean {\teff} surface at much smaller $\tau$.

The simulations both produce plausible {\ttau} relations.  While distinct from each other, they are both within the range of solar semi-empirical {\ttau} relations.  Both simulations are closest to the analytic Eddington relation at optical depths near to or greater than unity, but deviate from it higher in the atmosphere.  At small optical depth ($\tau < 0.2$) the ray simulation is closest to the relation of \citet{1981ApJS...45..635V}, while the 3D Eddington simulation appears to be similar to \citet{1966ApJ...145..174K}.

\subsection{Turbulent Pressure and Gas Dynamics}

The convective velocities in the SAL result in added support in the form of turbulent pressure.  Stellar models computed with mixing length theory do not include the effects of turbulent pressure. The added pressure from turbulence modifies the hydrostatic balance and is an important consideration for improving stellar models.

\begin{figure}
\epsscale{1.0}
\plotone{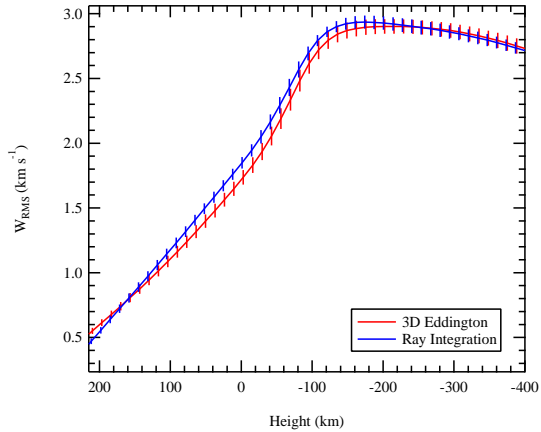}
\caption{The RMS of the vertical velocity remains essentially unchanged. Changing the radiative transfer scheme introduced only very small differences above the SAL.   The surface is defined as height at which $<T>=<T_{\rm eff}>$, and the peak of the SAL is approximately 80 km below the surface.}
\label{fig:wrms}
\end{figure}

Velocity statistics remain largely unchanged between the simulations.  Figure \ref{fig:wrms} shows that the RMS of the vertical velocity as a function of height is nearly the same in both the ray and 3D Eddington simulations.  The RMS vertical velocities are identical below the peak of the SAL, and show very minor differences above.  This is because the velocities are generated in the adiabatic convectively unstable region, and both simulations have the same adiabat.  As seen in Figure \ref{fig:wrms}, they can only start to deviate from each other in the SAL and convective overshooting regions where the different radiative transfer schemes have altered the stratification.

Convection provides a significant source of support in the SAL.  Turbulent pressure is defined as:
\begin{equation}
P_{\rm turb}=\rho w_{\rm rms}^2,
\end{equation}
where $w_{\rm rms}$ is the RMS of the vertical velocity. Turbulent pressure is small in the adiabatic region where gas pressure is high, reaches a maximum in the SAL, and drops quickly in the atmosphere where gas density is low.  The two simulations show a relatively small change in {\pturbopgas} from 13\% to 14\% when the radiative transfer model is changed.  Again, this is primarily a result of the differences in thermal structure (the density is different for a given temperature) since we see little change in the RMS velocity. 

\begin{figure}
\epsscale{1.0}
\plotone{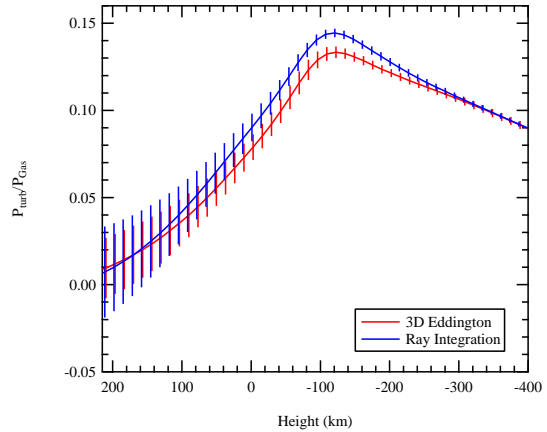}
\caption{Turbulent Pressure as a function of height.  Changing the radiative transfer scheme results in a small change in the peak turbulent pressure.  Changes in turbulent pressure are primarily a result of the different mean stratifications, and not changes in the gas dynamics.  The surface is defined as height at which $<T>=<T_{\rm eff}>$, and the peak of the SAL is approximately 80 km below the surface.}
\label{fig:pturb}
\end{figure}

While the radiative transfer scheme is the sole cause for the differences in Figure \ref{fig:pturb}, turbulent pressure is also sensitive to other aspects of the simulation.  As was the case with the mean stratification in Figure \ref{fig:rho}, the  change in turbulent pressure caused by the different radiative transfer schemes is within the range of what is seen in simulations computed using different codes.  For example, \citet{2012A&A...539A.121B} shows {\pturbopgas} ranging from 16\% to 18\% between solar simulations computed with the STAGGER, CO5BOLD, and MURaM codes.

\section{Conclusions}

We used 3D radiation hydrodynamic simulations to compute the thermal structure near the surface of stars where energy transport transitions from fully convective to fully radiative.  Since convection is multidimensional, achieving proper domain coverage is essential in providing accurate radiative transfer.  Radiative transport in 3D is computationally demanding, and to keep computational costs manageable, we make approximations to either the transfer equation or to domain coverage.

Our convection code includes two radiative transfer schemes that fundamentally differ in their methods for domain coverage.  The commonly used long-characteristic ray integration provides accurate solutions along individual rays, but lack of computational power necessitates under-sampling the domain and precludes proper 3D coverage.  The highly variable and asymmetric nature of {\qrad} makes it difficult to accurately sample the full 3D domain unless a great many rays are used, which is currently computationally cost prohibitive.  Our experience with undersampled ray integration suggests that it is important to consider the placement of the rays with respect to the highly asymmetric local flux, as the total flux may be sensitive to how well the local flux is sampled by the rays. We contrast the ray method with the 3D Eddington approximation, which is derived from moment equations closed at the level of first order, and has some computational advantages over long characteristic ray integration.

We find good general agreement in mean quantities from simulations computed with the 3D Eddington approximation and the long-characteristic ray integration method.  Both simulations have the same adiabatic structure in the convective interior, and the same energy flux.  They do, however, differ in radiative efficiency as a function of depth.  The differences in energy transport primarily affects the thermal structure of the SAL and atmosphere, while leaving the gas dynamics largely unchanged.

The most significant differences we find are in the mean temperature stratification through the SAL and the {\ttau} relation in the atmosphere. Although the {\ttau} relations are distinct, they both compare well with semi-empirical solar {\ttau} relations.  Limb-darkening measurements may lead to improved constraints on the solar {\ttau} relation.  We note that results from ray integration and the 3D Eddington approximation may differ in the photosphere beyond the domain of the simulations studied here.  Further study is necessary to determine whether there are differences in the photosphere of the 3D RHD simulations, and to what extent any differences have on observable quantities, such as features derived from spectral line synthesis using structures computed from the 3D simulations.

Simulations computed with the 3D Eddington approximation and under-sampled long-characteristic ray integration both provide similar thermal structures that self-consistently couple the region of efficient adiabatic convection to the atmosphere.  Indeed, the differences seen by changing the radiative transfer scheme are within the range of results of solar simulations computed using different codes \citep{2009MmSAI..80..701K,2012A&A...539A.121B}.  We therefore conclude that both radiative transfer techniques give rise to robust results for the angular resolution afforded by current computational power.

\acknowledgments

The authors thank the anonymous referee for insightful comments and suggestions that have helped to improve the paper. JT is supported by a PGS-D scholarship of the Natural Sciences and Engineering Research Council of Canada and by NASA ATFP grant \#NNX09AJ53G to SB. This work was also supported in part by the facilities and staff of the Yale University Faculty of Arts and Sciences High Performance Computing Center.

%\bibliography{references}

\end{document}